\documentclass[prb,reprint,ttisuperscriptaddress,twocolumn,10pt]{revtex4-1}
\usepackage[dvips]{graphicx}
\usepackage{subfigure}
\usepackage{bm}
\usepackage{color}
\usepackage{amsmath,amssymb}
\begin{document}
\title{
  Spin-motive force due to domain wall motion in the presence of Dzyaloshinskii-Moriya Interaction 
}
\author{Yuta Yamane}
\affiliation{Center for Emergent Matter Science (CEMS), RIKEN, Wako, Saitama 351-0198, Japan}
\date{\today}
\begin{abstract}
We theoretically demonstrate that the presence of Dzyaloshinskii-Moriya Interaction (DMI) can lead to enhancement of the spin-motive force (SMF) arising due to field-induced ferromagnetic domain wall motion.
A SMF refers to an electric voltage induced by dynamical magnetic textures, which reflects the temporal and spatial variations of the magnetization.
A DMI can introduce extra spatial rotation of the magnetization in the domain wall region, which turns out to cause the enhancement of the SMF.
We derive an expression for the SMF, and examine the field- and DMI-dependences of the SMF.
We find that the SMF can be amplified by up to an order of magnitude in the low field regime, where the external field is lower than the so-called Walker breakdown field.
\end{abstract}
\maketitle
\newpage
The exchange interaction between the conduction electron spin and the local magnetization in magnetic materials is responsible for a variety of important phenomena.
Among the spintronic effects caused by this interaction, spin-transfer torque\cite{Slonczewski,Berger-stt} paves a path to promising information technology, providing an efficient way of manipulating the magnetization by charge current.\cite{Brataas}
The same interaction can also mediate an electric-voltage generation by dynamical magnetic textures.
This electric voltage (or the mechanism that induce the electric voltage) is known as spin-motive force (SMF).\cite{Berger,Stern,Barnes,Ieda,Volovik2013,Hals}
A SMF reflects temporal and spatial variations of the magnetization, and thus offers a powerful method to probe and explore various dynamical magnetic textures, such as a moving domain wall (DW),\cite{Berger,Barnes,Stamenova,Zhang2009,Zhang2010,Yang,Kim2011,Hayashi} magnetic vortex,\cite{Ohe,Moon,Tanabe} and skyrmion lattice.\cite{Shimada,Yamane2014}

Theoretically, SMF can be attributed to a spin-dependent electric field,\cite{Korenman,Volovik,Yamane2011-jap} which is often referred to as spin electric field, arising due to the exchange coupling and acting on the conduction electrons.
While the basic concept and theoretical framework of SMF had been established through 1970s-90s,\cite{Korenman,Berger,Volovik,Stern,Aharonov,Ryu} the first experimental confirmations had to wait until the late 2000s\cite{Yang,Hai,Yamane2011-prl,Hayashi,Tanabe,Nagata} since it requires a control of dynamical magnetic textures at the precision of submicron meter scales.
The development of the SMF theory in the past decade has shedded light on the roles of the nonadiabaticity in electron spin dynamics\cite{Saslow,Duine,Tserkovnyak,Shibata,Lucassen} and the Rashba spin-orbit coupling\cite{Jalil,Kim,Tatara,Yamane2013,Ho} on the SMF.
The possibility of SMF in antiferromagnetic materials has recently been pointed out.\cite{Cheng,Gomonay,Okabayashi,Yamane2016}

The experimentally observed SMFs thus far are typically 100 nV - 1 $\mu$V in magnitude.\cite{Yang,Yamane2011-prl,Hayashi,Tanabe,Nagata}
To achieve larger SMF is deemed indispensable towards realization of spintronic devices actively exploiting SMF. 
In this article we address this problem demonstrating that, in the presence of Dzyaloshinskii-Moriya Interaction (DMI), the SMF due to field-induced DW motion can be enhanced by up to an order of magnitude in low field regime.\cite{Note}
A DMI arises in systems with broken inversion symmetry,\cite{Dzyaloshinsky,Moriya} favoring spatially rotating magnetic structures with a specific rotational sense.
In the present study, we focus on the so-called bulk DMI,\cite{Muhlbauer,Yu,Yu2011,Seki} which emerges due to noncentrosymmetric crystal structures such as in B20 compounds.\cite{Pfleiderer}
The presence of bulk DMI leads to extra spatial rotation of the magnetization in the DW region,\cite{Tretiakov,Kravchuk,Wang,Zhuo} which turns out to play a crucial role in the enhancement of the SMF.
We derive an expression for the SMF, and examine the field- and DMI-dependences of the SMF.
Our results suggest a new perspective on DMI materials as an suitable stage for pursuit of larger SMF and for certain types of SMF applications.

\emph{Domain wall dynamics ---}
Let us begin by examining the field-induced DW dynamics in the presence of DMI.
We consider a one-dimensional ferromagnetic nanowire extending along the $z$ axis (the inset of Fig.~1), whose magnetic energy density $u$ is assumed as
\begin{eqnarray}
  u &=& A \left(\frac{\partial{\vec m}}{\partial z}\right)^2 - K m_z^2 
  + K_\perp m_y^2  \nonumber \\ &&
  - D \left(m_x\frac{\partial m_y}{\partial z} - m_y \frac{\partial m_x}{\partial z}\right)
  - \mu_0 M_{\rm S} {\vec m} \cdot {\vec H}  ,
\end{eqnarray}
where ${\vec m}$ is the unit vector that defines the magnetization direction, $A$ is the exchange stiffness, $K(>0)$ and $K_\perp(>0)$ are the easy-axis and hard-axis anisotropy constants, respectively, $D$ is the DMI constant, $M_{\rm S}$ is the saturation magnetization, and ${\vec H}$ is the external magnetic field.
Our form of DMI corresponds to the so-called Dzyaloshinskii vector lying in the $z$ axis, and here we assume $D>0$.

In the parameter regime of $D^2>4AK$ the ground state is unique, which is a magnetic spiral configuration, and it prevents the formation of a DW.\cite{Tretiakov}
For $D^2<4AK$, on the other hand, the two solutions $m_z = \pm1$ minimize the magnetic energy, thus allowing a DW to exist as a transition region from one solution to another.
Assuming the hard-axis anisotropy to be small compared to the easy-axis anisotropy and the DMI, an equilibrium DW solution that locally minimizes the magnetic energy is given by\cite{Tretiakov,Kravchuk}
\begin{eqnarray}
  \theta\left(z\right)  &=&  2 \tan^{-1} e^{Q\left(z - z_0\right) / \Delta}  , \label{theta} \\
  \phi\left(z\right)  &=&  \Gamma z + \varphi  , \label{phi}
\end{eqnarray}
where the polar angles $(\theta,\phi)$ are defined by ${\vec m} = (\sin\theta\cos\phi,\sin\theta\sin\phi,\cos\theta)$, $z_0$ represents the DW center position, $\Delta$ is the DW width parameter given by $\Delta = \Delta_0 (1-D^2/4AK)^{-1/2}$ with $\Delta_0 = (A/K)^{1/2}$, $Q$ is the topological charge of the DW defined by $Q = \pi^{-1} \int_{-\infty}^\infty dz (\partial\theta/\partial z)= \pm1$ ($Q=+1$ corresponds to a head-to-head DW, while $Q=-1$ to a tail-to-tail one), $\Gamma = D/2A$, and $\varphi$ is a constant that takes $0$ or $\pi$ in eqilibrium.
In the absence of DMI ($D=0$), Eqs.~(\ref{theta}) and (\ref{phi}) reduce to the usual Walker solution with $\Delta=\Delta_0$ and $\Gamma=0$.\cite{textbook}

The DW can be driven into motion by an external magnetic field ${\vec H} = H {\vec e}_z$.
We here assume $H$ to be sufficiently weak that the dynamical DW sustains the structure of Eqs.~(\ref{theta}) and (\ref{phi}), but with $z_0(t)$ and $\varphi(t)$ becoming time dependent.
In this case, since the time evolution of ${\vec m}$ occurs only through that of $z_0$ and $\varphi$, these two parameters are regarded as the collective coordinates for the DW dynamics\cite{textbook};
the variation of $z_0(t)$ corresponds to the translational motion of the DW along the nanowire, while $\varphi(t)$ describes the rotational motion of the DW magnetization around the $z$ axis.

The dynamics of the magnetization ${\vec m}$ in general obeys the Landau-Lifshitz-Gilbert equation
\begin{equation}
  \frac{\partial{\vec m}}{\partial t}  =  - \gamma {\vec m} \times {\vec H}_{\rm eff}
                                                        + \alpha {\vec m} \times \frac{\partial{\vec m}}{\partial t}  ,
\label{llg}
\end{equation}
where $\gamma$ is the gyromagnetic ratio, $\alpha$ is the Gilbert damping constant, and ${\vec H}_{\rm eff} = - (\mu_0 M_{\rm S})^{-1} \delta u / \delta{\vec m}$ is the effective magnetic field.
Eq.~(\ref{llg}) with the above-introduced ansatz leads to a set of equations of motion for $(z_0,\varphi)$,\cite{Zhuo}
\begin{eqnarray}
  \frac{dz_0}{dt}  &=& \frac{Q\gamma\Delta}{1+\alpha^2}
                                   \left[ \alpha  H 
                                           + \frac{\zeta\left(1 - Q\alpha\Gamma\Delta\right)}{\Delta}
                                              \frac{H_k}{2} \sin2\varphi
                                   \right]  ,  \label{dz0} \\
  \frac{d\varphi}{dt}  &=& \frac{\gamma}{1+\alpha^2}
                                        \left[ \left(1 + Q\alpha\Gamma\Delta\right) H
                                                - \frac{\zeta\Delta}{\Delta_0^2} \frac{\alpha H_k}{2} \sin2\varphi
                                        \right]  ,  \label{dphi}
\end{eqnarray}
where $H_k = 2K_\perp / \mu_0M_{\rm S}$ and $\zeta = \pi \Gamma\Delta^2 / \sinh\left(\pi\Gamma\Delta\right)$.

Fig.~1 plots the DW velocity $v = T^{-1} \int_0^T dt (dz_0/dt)$, obtained by numerically simulating Eqs.~(\ref{dz0}) and (\ref{dphi}) from $t=0$ to $T=10$ $\mu$s, as a function of $|H|$ for four different sets of  ($D,Q$).
Cases I, II, III, and IV correspond to ($10^{-3}$ Jm$^{-2}$,+1), ($10^{-3}$ Jm$^{-2}$,$-1$), (0,+1), and (0,$-1$), respectively.
The other parameters are common for the four cases, which are:
$A=10^{-11}$ Am$^{-1}$, $K=4\times10^5$ Jm$^{-3}$, $K_\perp=10^5$ Jm$^{-3}$, $M_{\rm S}=6\times10^5$ Am$^{-1}$, $\gamma=2.211\times10^5$ (Am$^{-1}$)$^{-1}$s$^{-1}$, and $\alpha=0.01$.
Notice that $H = - |H|$, as depicted in the inset of Fig.~1.
The DW mobility $|\partial v/ \partial H|$ sharply drops at the so-called Walker breakdown field\cite{Schryer} $H_{\rm W}$, which is in the presence of bulk DMI given by\cite{Zhuo}
\begin{equation}
  H_{\rm W}  =  \frac{\zeta\Delta / \Delta_0^2}{1 + Q\alpha\Gamma\Delta} \frac{\alpha H_k}{2} ,
\label{Hw}
\end{equation}
and estimated as $\simeq16$ Oe for I, $\simeq15.9$ Oe for II, and $\simeq16.7$ Oe for III and IV, respectively. 
For $|H|<H_{\rm W}$, the last term in Eq.~(\ref{dphi}) cancels out the other terms at $\varphi = \frac{1}{2} \sin^{-1}\frac{2 \left(1 + Q\alpha\Gamma\Delta\right)}{\alpha} \frac{\Delta_0^2}{\zeta\Delta}\frac{H}{H_k}$, resulting in a purely translational DW motion with $d\varphi/dt=0$.
Once $|H|$ exceeds $H_{\rm W}$, the rotational dynamics with $d\varphi/dt\neq0$ takes place, leading to the decrease in the DW mobility.

\begin{figure}
  \centering
  \includegraphics[width=8cm,bb=0 0 865 639]{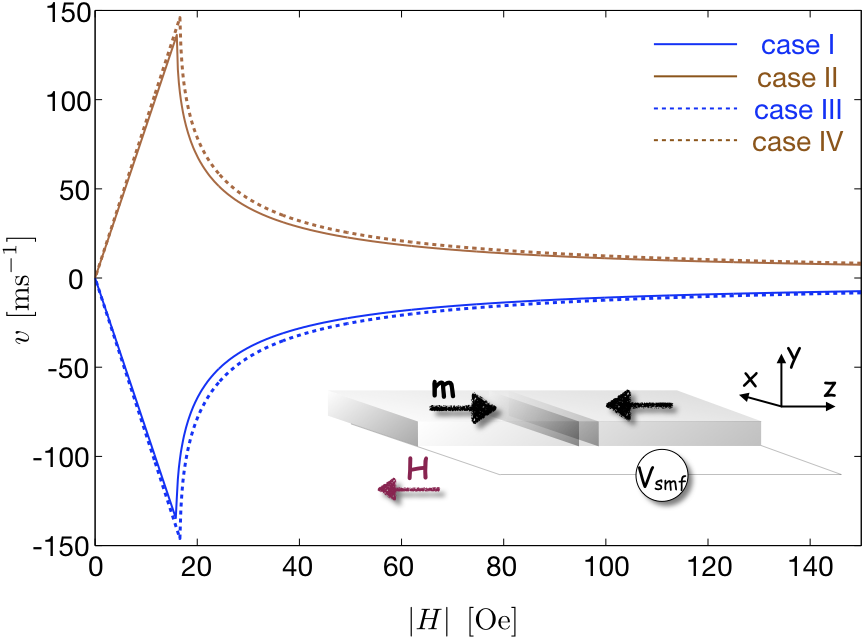}
  \caption{ Magnetic field $|H|$ dependence of DW velocity $v$ for four different sets of ($D,Q$).
                 Cases I, II, III, and IV show the results for ($10^{-3}$ Jm$^{-2}$,+1), ($10^{-3}$ Jm$^{-2}$,$-1$), (0,+1), and (0,$-1$), respectively;
                 the color of the curves indicates the sign of $Q$ (blue for $Q+1$ and brown for $Q=-1$), while the style of the curves distinguishes $D\neq0$ (solid curves) and $D=0$ (dashed curves).
                 It is seen that the DMI shifts the Walker breakdown fields $H_{\rm W}$ by several percent.
                 In the inset, schematic of the studied system is depicted.
               }
  \label{fig01}
\end{figure}

Within the collective-coordinate model with our present choice of parameter values, which are in a reasonable range for typical bulk-DMI materials,\cite{Kim2018} the effect of the DMI on DW velocity is merely to reduce $H_{\rm W}$ by several percent.
We will find shortly, nevertheless, that the DMI can have a major impact on the SMF that is induced by the DW motion.
For more systematic study of the field-driven DW dynamics itself in the presence of bulk DMI, see Ref.~\cite{Zhuo}, where the collective-coordinate model is compared with micromagnetic simulations.

\emph{Electric voltage generation ---}
Now let us discuss the SMF induced by the DW dynamics.
In an itinerant ferromagnet, the conduction electrons are subject to the spin electric field\cite{Korenman,Volovik,Saslow,Tserkovnyak,Duine,Yamane2011-jap,Shibata}
\begin{eqnarray}
  {\cal E}  &=&  \frac{P\hbar}{2e}
                        \left[ \sin\theta \left( \frac{\partial\theta}{\partial t} \frac{\partial\phi}{\partial z} 
                                                       - \frac{\partial\theta}{\partial z} \frac{\partial\phi}{\partial t} 
                                               \right)  \right. \nonumber \\ && \left.
                                + \beta \left( \frac{\partial\theta}{\partial t} \frac{\partial\theta}{\partial z} 
                                                    + \sin^2\theta \frac{\partial\phi}{\partial t} \frac{\partial\phi}{\partial z}
                                            \right)
                        \right]  ,
\label{e}
\end{eqnarray}
which arises as a result of the electron-magnetization exchange interaction.
Here $P$ represents the spin polarization of the conduction electrons, and $\beta$ is the dimensionless parameter characterizing the nonadiabaticity in the electron spin dynamics.
Eq.~(\ref{e}) requires the temporal and spatial derivatives of ${\vec m}$ to be finite simultaneously, and this condition is indeed satisfied around the dynamical DW.

The spin electric field can accelerate the conduction electrons in the same fashion as the ordinary electric field does, resulting in the electric voltage $V_{\rm smf} = \int_{-\infty}^\infty dz\ {\cal E} $ appearing across the DW.
In the absence of DMI, $\phi$ is spatially uniform (see Eq.~(\ref{phi})) and thus the terms that contain $\partial\phi/\partial z$ in Eq.~(\ref{e}) vanish.
When $D\neq0$, in contrast, these terms can no longer be ignored since $\partial\phi/\partial z= \Gamma$.
We will show that the $\frac{\partial\theta}{\partial t}\frac{\partial\phi}{\partial z}$ term in Eq.~(\ref{e}) indeed provides the most dominant contribution to $V_{\rm smf}$ in the field regime of $|H|<H_{\rm W}$.

Using the DW dynamics obtained by the collective-coordinate approach and doing some elementary algebra, one obtains
\begin{equation}
  V_{\rm smf}  =  - \frac{P\hbar}{e} \left[ \left(\beta + Q\Gamma\Delta\right) \frac{1}{\Delta}\frac{dz_0}{dt}
                                              + \left( Q - \beta\Gamma\Delta \right) \frac{d\varphi}{dt} 
                                      \right]  .              
\label{v}
\end{equation}
Eq.~(\ref{v}) contains our central results, revealing the way the DMI contributes to the SMF.
For $D=0$, Eq.~(\ref{v}) reproduces the expression for the SMF known from the previous studies.\cite{Duine,Lucassen}
In the following, we examine Eq.~(\ref{v}) more closely.

\begin{figure}
  \centering
  \includegraphics[width=8cm,bb=0 0 861 663]{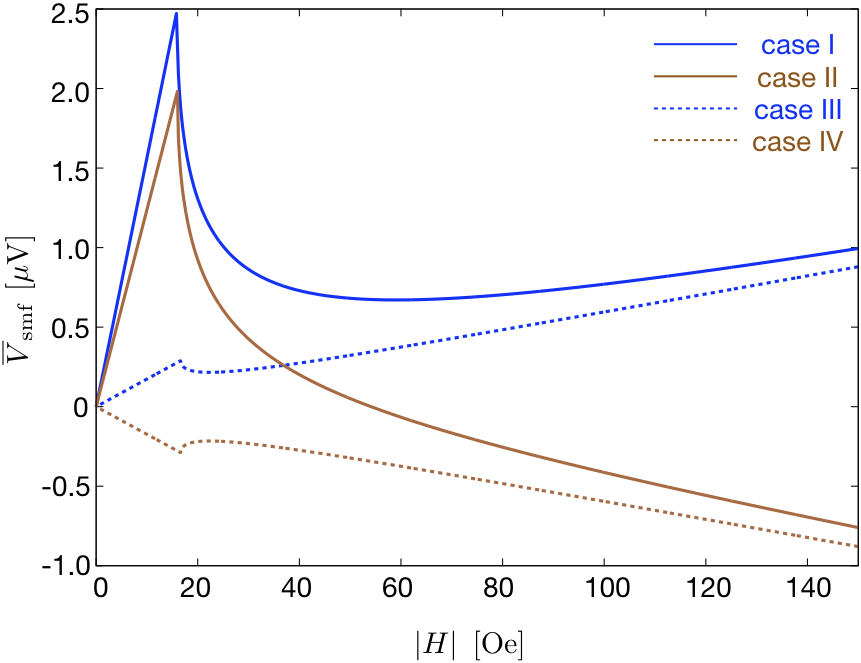}
  \caption{ Magnetic field $|H|$ dependence of the time-averaged SMF $\overline{V}_{\rm smf}$.
                 Cases I, II, III, and IV refer to the same sets of ($D,Q$) as in Fig.~1.
                 The effect of the DMI is most pronounced for $|H|<H_{\rm W}\simeq16\sim17$ Oe, where $|\overline{V}_{\rm smf}|$ for $D\neq0$ are nearly an order of magnitude greater than those for $D=0$.
                 As $|H|$ is increased passing $H_{\rm W}$, the $\overline{V}_{\rm smf}$ curves with $D\neq0$ converge to the curves for $D=0$.
                 }
\end{figure}

Fig.~2 displays the time-averaged SMF $\overline{V}_{\rm smf} = T^{-1} \int_0^T dt V_{\rm smf} (t)$ as a function of $|H|$, where $V_{\rm smf}(t)$ is computed by Eq.~(\ref{v}), with $dz_0/dt$ and $d\varphi/dt$ numerically simulated as before.
Cases I-IV refer to the same sets of parameters as in Fig.~1.
For $P$ and $\beta$, we employed $P=0.5$ and $\beta=0.03$ for all the four cases.

\begin{figure}
  \centering
  \includegraphics[width=8cm,bb=0 0 879 655]{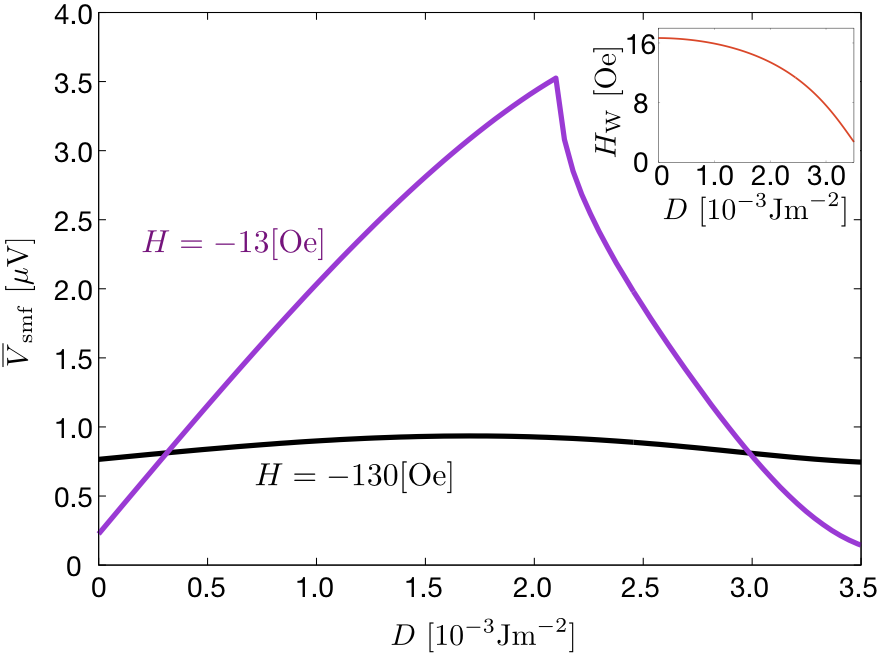}
  \caption{ DMI constant $D$ dependences of the time-averaged SMF $\overline{V}_{\rm smf}$, with $H$ fixed at $H=-13$ Oe (purple curve) and $H=-130$ Oe (black curve).
  In the latter case, $\overline{V}_{\rm smf}$ is relatively insensitive to $D$ because $|H|\gg H_{\rm W}$ at any value of $D$.
  For the case with $H=-13$ Oe, on the other hand, $\overline{V}_{\rm smf}$ increases with $D$ up to $D=D_{\rm c}\simeq2.1\times10^{-3}$ Jm$^{-2}$, while decreases with $D$ for $D>D_{\rm c}$.
  This reflects the fact that at $D=D_{\rm c}$, $|H|$ coincides with $H_{\rm W}$.
 The other parameters are taken as the same in the previous calculations, and $Q=+1$ for both cases.
                 In the inset, $D$ dependence of $H_{\rm W}$ is shown, where $H_{\rm W}$ monotonically decreases as $D$ increases, and converges to zero as $D\rightarrow(4AK)^{1/2}=4\times10^{-3}$ Jm$^{-2}$.
                 }
\end{figure}

The influence of the DMI is most prominent for $|H|<H_{\rm W}\simeq16\sim17$ Oe, where $\overline{V}_{\rm smf}$ exhibits the linear dependence on $|H|$ for all the four cases, while its slope is remarkably enhanced in the presence of DMI.
As for the sign of $\overline{V}_{\rm smf}$, it is positive for both I and II regardless of the sign of $Q$.
This is in contrast to the simple linear $Q$-dependence for $D=0$, i.e., $\overline{V}_{\rm smf}(H,D=0,Q=+1) = - \overline{V}_{\rm smf}(H,D=0,Q=-1)$.
In this field regime, an analytical expression for the SMF is available from Eqs.~(\ref{dz0}), (\ref{dphi}), and (\ref{v}) as
\begin{equation}
  V_{\rm smf}  =  - \frac{P\hbar}{e} \frac{Q\beta + \Gamma\Delta}{\alpha} \frac{\Delta_0^2}{\Delta^2} \gamma H  ,\qquad
  (|H|<H_{\rm W}) .
\label{v2}
\end{equation}
This is time independent, and can be directly compared to $\overline{V}_{\rm smf}$ in Fig.~2.
Because $\Gamma\Delta \simeq 0.55 \gg \beta$, the $Q$-independent $\Gamma\Delta$ term in Eq.~(\ref{v2}) dominates the other one, which explains the above-mentioned behavior of $\overline{V}_{\rm smf}$.
Notice that $\overline{V}_{\rm smf}$ reaches as high as $\sim2.5$ $\mu$V at $|H|\simeq H_{\rm W}\simeq16$ Oe for case I.
The largest experimentally-observed SMF due to DW motion thus far is $\sim1$ $\mu$V with $|H|\simeq150$ Oe, exploiting Permalloy nanowires.\cite{Hayashi}

As $|H|$ is increased passing $H_{\rm W}$, the impact of the DMI on the SMF diminishes;
after hitting the peaks at $|H|=H_{\rm W}$, the $\overline{V}_{\rm smf}$ curves for $D\neq0$ sharply plunge and approach the curves for $D=0$ (I approaches to III, and II to IV, as seen in Fig.~2).
This may be understood from the fact that, in Eq.~(\ref{v}), the $dz_0/dt$ term is dominated by the DMI contribution since $\Gamma\Delta\gg\beta$, while the DMI is less important for the $d\varphi/dt$ term because $1\gg \beta\Gamma\Delta$.
For $|H|<H_{\rm W}$, the DW dynamics is a pure translational morion ($d\varphi/dt=0$), and this is because in this field regime the effect of the DMI is most pronounced, as discussed before.
For $|H|> H_{\rm W}$, the decrease in $dz_0/dt$ and the switching on of $d\varphi/dt$ spoil the influence of the DMI.
When $|H|$ is large enough compared to $H_{\rm W}$ so that the oscillating terms ($\propto\sin2\varphi$) in Eqs.~(\ref{dz0}) and (\ref{dphi}) can be neglected in the time-averaged DW dynamics, an approximate expression for the SMF may be given by
\begin{equation}
  \overline{V}_{\rm smf}  \simeq  - \frac{P\hbar}{e}
                                     \left[Q - \Gamma\Delta \left(\beta - 2\alpha\right)\right] \gamma H  , \qquad
                                     (|H| \gg H_{\rm W})  ,
\label{v3}
\end{equation}
where $\alpha,\beta\ll1$ has been used.
Since $\Gamma\Delta(\beta-2\alpha)\ll1$, the DMI only presents a small correction to $\overline{V}_{\rm smf}$ in this field regime, being consistent with the above argument.

We should remark here that a larger DMI does not always lead to a higher SMF because, according to Eq,~(\ref{Hw}), $H_{\rm W}$ monotonically decreases as $D$ increases, see the inset of Fig.~3.
We plot the $D$-dependence of $\overline{V}_{\rm smf}$ in Fig.~3, where the purple and black curves show the results with $H$ being fixed at $-13$ Oe and $-130$ Oe, respectively, and $Q=+1$ for both cases.
(The other parameters are again the same as before.)
For the latter case, $|H|$ is well above $H_{\rm W}$ regardless of the value of $D$, and the SMF is thus relatively insensitive to $D$ as discussed before. 
For $H=-13$ Oe, on the other hand, $|H|<H_{\rm W}$ for $D<D_{\rm c}\simeq2.1\times10^3$ Jm$^{-2}$, while $|H|>H_{\rm W}$ for $D>D_{\rm c}$.
This is why $\overline{V}_{\rm smf}$ increases with $D$ up to $D_{\rm c}$, and decreases with $D$ for $D>D_{\rm c}$. 
Desirable materials, in terms of pursuing larger SMF, would have large DMI as well as large magnetic anisotropies.

In conclusion, we have theoretically demonstrated that DMI is capable of dramatically amplifying the SMF arising due to field-induced DW dynamics. 
Importance of DMIs in the magnetism community has been growing in recent years since they can stabilize magnetic skyrmion lattices as well as individual skyrmions, which exhibit various characteristic properties that are advantageous for technological applications.\cite{Fert,Nagaosa,Finocchio,Fert2017}
In the context of DW physics, a surface DMI in perpendicularly-magnetized materials renders a DW a specific chirality,\cite{Heide,Thiaville} leading to high efficiency in current-driven DW motion.\cite{Emori}
We believe our results have revealed a novel importance of bulk DMI in the SMF-related DW physics, and have made a vital step towards realization of SMF-based spintronic devices.

The author appreciates Dr. J. Ieda for valuable comments on the manuscript.
This research was supported by Research Fellowship for Young Scientists from Japan Society for the Promotion of Science.



\begin{thebibliography}{99}
\bibitem{Slonczewski}
    J. C. Slonczewski, J. Magn. Magn. Matter {\bf159}, L1 (1996).
\bibitem{Berger-stt}
    L. Berger, Phys. Rev. B {\bf54}, 9353 (1996).
\bibitem{Brataas}
   A. Brataas, A. D. Kent, and H. Ohno, Nat. Mater. {\bf11}, 372 (2012).
\bibitem{Berger}
    L. Berger, Phys. Reb. B {\bf 33}, 1572 (1986).
\bibitem{Stern}
    A. Stern, Phys. Rev. Lett. {\bf 68}, 1022 (1992).
\bibitem{Barnes}
    S. E. Barnes and S. Maekawa, Phys. Rev. Lett. {\bf 98}, 246601 (2007).
\bibitem{Ieda}
    J. Ieda, Y. Yamane, and S. Maekawa, SPIN {\bf 03}, 1330004 (2013).
\bibitem{Volovik2013}
  G. E. Volovik, JETP Letters {\bf98}, 480 (2013).
\bibitem{Hals}
    K. M. D. Hals and A. Brataas, Phys. Rev. B {\bf 91}, 214401 (2015).
\bibitem{Stamenova}
    M. Stamenova, T. N. Todorov, and S. Sanvito, Phys. Rev. B {\bf77}, 54439 (2008).
\bibitem{Yang}
    S. A. Yang, G. S. D. Beach, C. Knutson, D. Xiao, Q. Niu, M. Tsoi, and J. L. Erskine,
    Phys. Rev. Lett. {\bf 102}, 067201 (2009);
    S. A. Yang, G. S. D. Beach, C. Knutson, D. Xiao, Z. Zhang, M. Tsoi, Q. Niu, A. H.MacDonald, and J. L. Erskine,
    Phys. Rev. B {\bf 82}, 054410 (2010).
\bibitem{Zhang2009}
    S. Zhang and S. S.-L. Zhang, Phys. Rev. Lett. {\bf102}, 086601 (2009).   
\bibitem{Zhang2010}
    S. S.-L. Zhang and S. Zhang, Phys. Rev. B {\bf82}, 184423 (2010). 
\bibitem{Kim2011}
    K.-W. Kim, J.-H. Moon, K.-J. Lee, and H.-W. Lee, Phys. Rev. B {\bf84}, 054462 (2011).
\bibitem{Hayashi}
    M. Hayashi, J. Ieda, Y. Yamane, J. I. Ohe, Y. K. Takahashi, S. Mitani, and S. Maekawa,
    Phys. Rev. Lett. {\bf 108}, 147202 (2012).
\bibitem{Ohe}    
    J. Ohe and S. Maekawa, J. Appl. Phys. {\bf 105}, 07C706 (2009); 
    J. Ohe, S. E. Barnes, H.-W. Lee, and S. Maekawa, Appl. Phys. Lett. {\bf 95}, 123110 (2009).
\bibitem{Moon}
    J.-H. Moon and K.-J. Lee, J. Magn. {\bf16}, 6 (2011).
\bibitem{Tanabe}
    K. Tanabe, D. Chiba, J. Ohe, S. Kasai, H. Kohno, S. E. Barnes, S. Maekawa, K. Kobayashi, and T. Ono,
    Nat. Commun. {\bf 3}, 845 (2012).
\bibitem{Shimada}
    J. Ohe and Y. Shimada, Appl. Phys. Lett. {\bf103}, 242403 (2013);
    Y. Shimada and J. Ohe, Phys. Rev. B {\bf91}, 174437 (2015).
\bibitem{Yamane2014}
    Y. Yamane, S. Hemmatiyan, J. Ieda, S. Maekawa, and J. Sinova, Sci. Rep. {\bf4}, 06901 (2014).
\bibitem{Yamane2011-jap}  
    Y. Yamane, J. Ieda, J. Ohe, S. E. Barnes, and S. Maekawa, J. Appl. Phys. {\bf 109}, 07C735 (2011).
\bibitem{Korenman}
    V. Korenman, J. L. Murray, and R. E. Prange, Phys. Rev. B {\bf 16}, 4032 (1977).
\bibitem{Volovik}
    G. E. Volovik, J. Phys. C {\bf 20}, L83 (1987).
\bibitem{Aharonov}
    Y. Aharonov and A. Stern, Phys. Rev. Lett. {\bf 69}, 3593 (1992).
\bibitem{Ryu}    
    C.-M. Ryu, Phys. Rev. Lett. {\bf 76}, 968 (1996).
\bibitem{Hai}
    P. N. Hai, S. Ohya, M. Tanaka, S. E. Barnes, and S. Maekawa, Nature (London) {\bf 458}, 489 (2009).
\bibitem{Yamane2011-prl}
    Y. Yamane, K. Sasage, T. An, K. Harii, J. Ohe, J. Ieda, S. E. Barnes, E. Saitoh, and S.Maekawa,
    Phys. Rev. Lett. {\bf 107}, 236602 (2011).
\bibitem{Nagata}
    M. Nagata, T. Moriyama, K. Tanabe, K. Tanaka, D. Chiba, J. Ohe, Y. Hisamatsu, T. Niizeki, H. Yanagihara, E. Kita, and T. Ono, Appl. Phys. Exp. {\bf 8}, 123001 (2015). 
\bibitem{Saslow}
   W. M. Saslow, Phys. Rev. B {\bf76}, 184434 (2007).
\bibitem{Duine}
    R. A. Duine, Phys.Rev.B {\bf 77}, 014409 (2008); {\it ibid.}, {\bf79}, 014407 (2009). 
\bibitem{Tserkovnyak}
    Y. Tserkovnyak and M. Mecklenburg, Phys. Rev. B {\bf 77}, 134407 (2008).
\bibitem{Shibata}
    J. Shibata and H. Kohno, Phys. Rev. B {\bf 84}, 184408 (2011).
\bibitem{Lucassen}
    M. E. Lucassen, G. C. F. L. Kruis, R. Lavrijsen, H. J. M. Swagten, B. Koopmans, and R. A. Duine,
    Phys. Rev. B {\bf84}, 014414 (2011).
\bibitem{Jalil}
    M. B. A. Jalil and S. G. Tan, IEEE Trans. Magn. {\bf46}, 1626 (2010).
\bibitem{Kim}
    K. W. Kim, J. H. Moon, K. J. Lee, and H. W. Lee, Phys. Rev. Lett. {\bf 108}, 217202 (2012).
\bibitem{Tatara}  
    G. Tatara, N. Nakabayashi, and K. J. Lee, Phys. Rev. B {\bf 87}, 054403 (2013).
\bibitem{Yamane2013}
    Y. Yamane, J. Ieda, and S. Maekawa, Phys. Rev. B {\bf 88}, 014430 (2013).
\bibitem{Ho}
    C. S. Ho, M. B. A. Jalil, and S. G. Tan, New J. Phys. {\bf17}, 123005 (2015).
\bibitem{Cheng}
    R. Cheng and Q. Niu, Phys. Rev. B {\bf 86}, 245118 (2012).
\bibitem{Gomonay}
    O. Gomonay, Phys. Rev. B {\bf91}, 144421 (2015).
\bibitem{Okabayashi}
    A. Okabayashi and T. Morinari, J. Phys. Soc. Jpn. {\bf 84}, 033706 (2015).
\bibitem{Yamane2016}
    Y. Yamane, J. Ieda, and J. Sinova, Phys. Rev. B {\bf 93}, 180408(R) (2016).
\bibitem{Note}
   By a ``low field'' we mean a magnetic field whose magnitude is smaller than the Walker breakdown field, which is introduced in Eq.~(\ref{Hw}).
\bibitem{Dzyaloshinsky}
    I. J. Dzyaloshinsky, Phys. Chem. Solids {\bf4}, 241 (1958).
\bibitem{Moriya}
    T. Moriya, Phys. Rev. {\bf120}, 91 (1960).
\bibitem{Muhlbauer}
    S. M\"{u}hlbauer, B. Binz, F. Jonietz, C. Pfleiderer, A. Rosch, A. Neubauer, R. Georgii, and P. Boni, Science {\bf323}, 915 (2009).
\bibitem{Yu}
    X. Z. Yu, Y. Onose, N. Kanazawa, J. H. Park, J. H. Han, Y. Matsui, N. Nagaosa, and Y. Tokura,
    Nature (London) {\bf465}, 901 (2010).
\bibitem{Yu2011}
    X. Z. Yu, N. Kanazawa, Y. Onose, K. Kimoto, W. Z. Zhang, S. Ishiwata, Y. Matsui, and Y. Tokura, Nat. Mater. {\bf10}, 106 (2011)
\bibitem{Seki}
    S. Seki, X. Yu, S. Ishiwata, and Y. Tokura, Science {\bf336}, 198 (2012).
\bibitem{Pfleiderer}
    C. Pfleiderer, T. Adams, A. Bauer, W. Biberacher, B. Binz, F. Birkelbach, P. B\"{o}ni, C. Franz, R. Georgii, M. Janoschek, F. Jonietz, T. Keller, R. Ritz, S. M\"{u}hlbauer, W. M\"{u}nzer, A. Neubauer, B. Pedersen, and A. Rosch, J. Phys.: Condens.Matter {\bf22}, 164207 (2010).
\bibitem{Tretiakov}
    O. A. Tretiakov and Ar. Abanov, Phys. Rev. Lett. {\bf 105}, 157201 (2010).
\bibitem{Kravchuk}
    V. P. Kravchuk, J. Magn. Magn, Mater. {\bf367}, 9 (2014).
\bibitem{Wang}
    W. Wang, M. Albert, M. Beg, M.-A. Bisotti, D. Chernyshenko, D. Cort\'{e}s-Ortu\~{n}o,
    I. Hawke, and H. Fangohr,
    Phys. Rev. Lett. {\bf114}, 087203 (2015).
\bibitem{Zhuo}
    F. Zhuo and Z. Z. Sun, Sci. Rep. {\bf 6}, 25122 (2016).
\bibitem{textbook}
    A. P. Malozemoff and J. C. Slonczewski, {\it Magnetic Domain Walls in Bubble Materials}
    (Academic, New York, 1979).
\bibitem{Schryer}
    N. L. Schryer and L. R. Walker, J Appl. Phys. {\bf45}, 5406 (1974).
\bibitem{Kim2018}
    K.-W. Kim, K.-W. Moon, N. Kerber, J. Nothhelfer, and K. Everschor-Sitte,
    Phys. Rev. B {\bf97}, 224427 (2018), and references therein.
\bibitem{Fert}
    A. Fert, V. Cros, and J. Sampaio, Nat. Nanotechnol. {\bf8}, 152 (2013).
\bibitem{Nagaosa}
    N. Nagaosa and Y. Tokura, Nat. Nanotechnol. {\bf8}, 899 (2013).
\bibitem{Finocchio}
    G. Finocchio, F. B\"{u}ttner, R. Tomasello, M. Carpentieri, and M. Kl\"{a}ui, J Phys. D: Condens. Matter {\bf49}, 423001 (2016).
\bibitem{Fert2017}
    A. Fert, N. Reyren, and V. Cros, Nat. Rev. Mater. {\bf2}, 17031 (2017).
\bibitem{Heide}
    M. Heide, G. Bihlmayer, and S. Bl\"{u}gel, Phys. Rev. B {\bf78}, 140403(R) (2008).
\bibitem{Thiaville}
    A. Thiaville, S. Rohart, E. Ju\'{e}, V. Cros, and A. Fert, Europhys. Lett. {\bf100}, 57002 (2012).
\bibitem{Emori}
    S. Emori, U. Bauer, S.-M. Ahn, E. Martinez, and G. S. D. Beach, Nat. Mater. {\bf12}, 611 (2013).
\end{thebibliography}
\end{document}